\documentclass[12pt,a4paper,final]{iopart}
\usepackage{iopams}  
\usepackage{graphicx}
\usepackage[breaklinks=true,colorlinks=true,linkcolor=blue,urlcolor=blue,citecolor=blue]{hyperref}

\begin{document}

\title{Towards high mobility InSb nanowire devices}

\author{\"Onder G\"ul$^{1*}$, David J.~van Woerkom$^{1*}$, Ilse van Weperen$^{1*}$, Diana Car$^2$, S\'ebastien R.~Plissard$^{1,2,3}$, Erik P.~A.~M.~Bakkers$^{1,2}$, Leo P.~Kouwenhoven$^1$}
\address{$^1$QuTech and Kavli Institute of Nanoscience, Delft University of Technology, 2600 GA Delft, Netherlands}
\address{$^2$Department of Applied Physics, Eindhoven University of Technology, 5600 MB Eindhoven, Netherlands}
\address{$^3$Present address: Laboratoire d'Analyse et d'Architecture des Syst\`emes, 7, avenue du Colonel Roche BP 54200 31031 Toulouse cedex 4, France}
\address{*Equal contribution}
\ead{\mailto{o.gul@tudelft.nl}}

\begin{abstract}
We study the low-temperature electron mobility of InSb nanowires. We extract the mobility at 4.2\,Kelvin by means of field effect transport measurements using a model consisting of a nanowire-transistor with contact resistances. This model enables an accurate extraction of device parameters, thereby allowing for a systematic study of the nanowire mobility. We identify factors affecting the mobility, and after optimization obtain a field effect mobility of $\sim2.5\mathbin{\times}10^4$\,cm$^2$/Vs. We further demonstrate the reproducibility of these mobility values which are among the highest reported for nanowires. Our investigations indicate that the mobility is currently limited by adsorption of molecules to the nanowire surface and/or the substrate.
\end{abstract}

\pacs{73.63.Nm, 81.05.Ea, 81.07.Gf, 81.16.-c, 85.30.Tv, 85.35.Be}
\vspace{2pc}
\noindent{\it Keywords}: field effect mobility, adsorption, nanowire transistor, nanowire FET, nanofabrication

\section{Introduction}

Advances in nanowire growth have led to development of novel quantum devices, such as cooper-pair splitters~\cite{Hofstetter2009}, hybrid semiconductor-superconductor devices~\cite{Doh2005} and spin-orbit qubits~\cite{Nadj-Perge2010}. Nanowire devices thus allow exploration of mesoscopic transport in a highly confined system and show potential as a quantum computation platform. Outstanding nanowire transport properties, such as a high level of tunability of device conductance and low disorder, have been essential to the realization of these experiments. 

Recently, hybrid superconductor-semiconducting nanowire devices have been identified~\cite{Lutchyn2010,Oreg2010} as a suitable platform to study Majorana end modes~\cite{Majorana1937}, zero-energy bound states that exhibit topological properties. Among various systems, InSb nanowires emerged as a very promising candidate due to their large spin-orbit interaction and large g factor. Reports on signatures of Majorana bound states in InSb nanowire-based systems followed quickly after their theoretical prediction~\cite{Mourik2012,Deng2012,Churchill2013}. To further develop this topological system, a reduction of the disorder in the nanowire is essential~\cite{Sau2012,Potter2011a}. Disorder reduces or even closes the topological gap that gives Majoranas their robustness, thereby impairing their use as topological qubits. Disorder is quantified by measurements of carrier mobility, which relates directly to the time between scattering events. Evaluation of carrier mobility in nanowires therefore indicates their potential for transport experiments and is thus crucial to further development of nanowire-based quantum devices.

According to the Matthiessen rule, various scattering mechanisms altogether determine the net mobility through~\cite{AshcroftMermin}
\begin{equation} \label{matt eq}
\frac{1}{\mu}=\frac{1}{\mu_1}+\frac{1}{\mu_2} + \ldots
\end{equation}
Here $\mu$ represents the net mobility which results from distinct scattering mechanisms each giving rise to a separate mobility $\mu_n$. In other words, the most dominant scattering contribution limits and hence determines the net mobility. Therefore the mobility can be improved by identifying the limiting mechanism and subsequently reducing or eliminating it.

Apart from the recently introduced Hall effect measurements on nanowires~\cite{Bloemers2012,Strom2012}, field effect transport measurements are the most common and experimentally most feasible method to extract charge carrier mobility in these systems. Here, one measures the current flowing through the nanowire channel contacted by two electrodes as a function of the gate voltage with fixed voltage bias. The conductance of the channel is described by the linear region of the accumulation regime of a field effect transistor (FET)~\cite{Sze1981}. In this case the conductance of the channel is
\begin{equation} \label{fet eq}
G(V_g)=\frac{\mu C}{L^2}\left(V_g-V_{th}\right),
\end{equation}
with gate voltage, $V_g$, mobility, $\mu$, capacitance, $C$, channel length, $L$, and threshold voltage, $V_{th}$. If the capacitance and the channel length are known, the field effect mobility can be determined from the transconductance, $g_m=dG/dV_g$. In most cases, to extract the mobility, the maximum (peak) transconductance is used. One should note that both the mobility and the field effect transport is described using the Drude model where charge carrier transport is classical and diffusive.

Previous studies showed that low-temperature field effect mobility for nominally undoped III-V nanowires is mainly limited by crystal defects such as stacking faults~\cite{Schroer2010,Kretinin2010,Shtrikman2011,Gupta2013a,Sourribes2014}, and surface effects such as surface roughness~\cite{Ford2012,Chuang2013}. Point defects are also thought to have an effect on the mobility~\cite{Bar-Sadan2012}. However, as they are difficult to detect so far no direct connection between impurities and mobility has been reported. Highest reported low-temperature field effect mobilities are 1.6 -- $2.5\mathbin{\times}10^4$\,cm$^2$/Vs. Such mobilities are observed in InAs nanowires~\cite{Schroer2010,Wang2013a}, InAs/InP core-shell nanowires~\cite{Jiang2007,VanTilburg2010} and GaN/AlN/AlGaN core-shell nanowires after correction for contact resistances~\cite{Li2006}. However, in most of these studies either data on a single device is reported, or the average mobility of several devices is significantly lower than the reported maximum~\cite{VanTilburg2010}. Systematic studies of such high-mobility nanowire FETs are thus largely lacking. 

Concerning field effect mobility, the InSb nanowires we investigate differ in several respects from their oft-studied InAs counterparts: the InSb nanowires we use have a larger diameter of approximately 100\,nm, reducing their surface-to-volume ratio compared to the thinner InAs nanowires, and are likely to have no surface accumulation layer. Instead, upward band bending leading to surface carrier depletion has been reported for both clean~\cite{Gobeli1965} and oxygen-covered InSb surfaces with (110) orientation, the orientation of our InSb nanowire facets. As the InSb facets are atomically flat no surface roughness is expected. Finally, the nanowires are purely zinc-blende and are free of stacking faults and dislocations. The growth of InSb nanowires we study is described in \cite{Plissard2012} and~\cite{Car2014}. Given the differences between InSb nanowires and other nanowire materials it is an open question what determines the low-temperature mobility in InSb nanowires. We note that while in~\cite{Plissard2012} field effect mobilities of these InSb wires are reported, no systematic investigation of the nanowire mobility was performed. The mobility extraction method presented here allows such a thorough investigation, thereby revealing new insights on nanowire mobility. 

To identify the factors affecting the mobility of InSb nanowires, we characterized the low-temperature mobility of nanowire FETs fabricated using different experimental parameters. We tailored the extraction of field effect mobility for the nanowires we study to accurately determine the essential transistor parameters of nanowire FETs. By systematic studies we developed a recipe that results in reproducible average mobilities of $\sim2.5\mathbin{\times}10^4$\,cm$^2$/Vs. While this value represents an average over many devices, the extracted mobility from a single measurement may exceed $3.5\mathbin{\times}10^4$\,cm$^2$/Vs. After optimizing the fabrication, we also find that adhesion of molecules to the nanowire and/or the substrate currently limits the extracted mobility. Although such adsorption effects are known to modify the nanowire conductance~\cite{Kretinin2010, Penchev2012} and also the room-temperature mobility~\cite{Dayeh2007,Du2009,Offermans2010} (note that ref.~\cite{Du2009} reports an increase of mobility upon adsorption, whereas ref.~\cite{Offermans2010} a reduction), our identification of surface adsorption being the limiting factor to low-temperature field effect mobility is new. The amount of adsorbates is reduced by evacuating the sample space for longer time prior to cool down and suggestions for further reduction of the adsorbates as well as to minimize their contribution to the field effect transport are made. We finally discuss various methods to investigate the surface properties of InSb nanowires.

\section{Experimental details}

\begin{figure}[!ht]
	\centering
 	\includegraphics[width=1\textwidth]{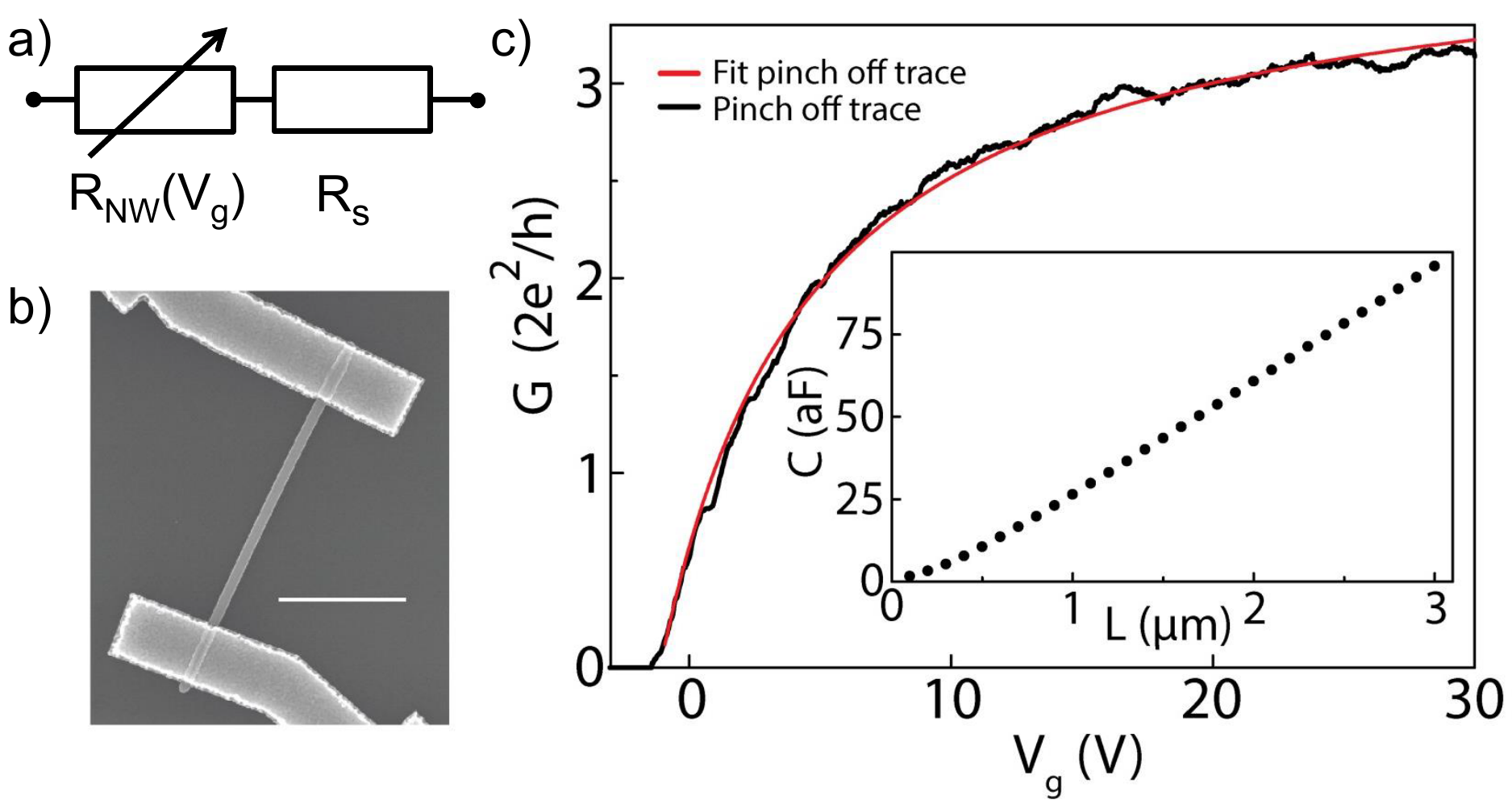}
 	\caption{\textbf{a)} Electrical diagram of the InSb nanowire FET. The FET is modelled as a nanowire channel with a resistance controlled by a nearby gate, $R_{NW}(V_g) = G_{NW}^{-1}(V_g)$, in series with fixed interface resistances, $R_s$. \textbf{b)} Electron microscope image of an InSb nanowire FET. Nanowire diameter is $\sim$~100\,nm. The nanowire is deposited onto a Si substrate covered with 285\,nm dry thermal SiO$_{2}$. Ti/Au (5/145\,nm) contacts have spacing of 1, 1.5, 2 or 2.5\,$\mu$m. Scale bar is 1\,$\mu$m. \textbf{c)} Conductance $G$, as a function of back gate voltage $V_g$ (black curve). Source-drain bias is set to 10\,mV throughout the study. Field effect mobility is extracted from a fit to the conductance (red curve) using eq.~\ref{mob_fit_expr}. All measurements are performed at a temperature of 4.2\,K. Inset: Gate-nanowire capacitance $C$, as a function of source-drain contact spacing $L$. Capacitance is extracted from a finite element model of the device geometry. Contacts are included in the simulated device geometry and lead to a non-linear $C$($L$) at small contact spacing.}
 	\label{MOB_fig1}
\end{figure}

InSb nanowire FETs are fabricated on a heavily doped Si substrate (used as a global back-gate) terminated with a 285-nm-thick dry thermal SiO$_2$ (Fig.~\ref{MOB_fig1}b). The substrate is patterned with alignment markers prior to nanowire deposition. Nanowires are positioned on the substrate using a micro-manipulator~\cite{Floehr2011}. Two terminal contacts are realized by electron beam lithography, metal evaporation (Ti/Au 5/145 nm) and lift-off. Argon plasma etching is employed prior to contact deposition. Further details about the fabrication process and the measurements can be found in Supplementary Text 1 and 2, respectively.

Due to the absence of a surface accumulation layer in InSb nanowires, an interface resistance of a few kilo ohms cannot be eliminated upon contacting the nanowire~\cite{vanWeperen2013}. Such interface resistances are known to reduce the transconductance, resulting in an underestimation of the intrinsic mobility~\cite{Schroder2006,Lu2008}. Moreover, at a temperature of 4\,K universal conductance fluctuations complicate the extraction of mobility from transconductance. We therefore tailor the extraction of field effect mobility to our InSb nanowire FETs~\cite{Plissard2013}. We model the interface resistances by a resistor $R_s$ with a fixed value (no gate voltage dependence), connected in series to the nanowire channel. A substantial part of the device resistance at high gate voltage stems from the interface resistances, strongly affecting the gate voltage dependent conductance. This complicates accounting for a possible change of mobility with gate voltage. We therefore assume a mobility independent of gate voltage. The device conductance is then given by (see also Fig.~\ref{MOB_fig1}a)
\begin{equation} \label{mob_fit_expr}
G(V_g)=\left(R_s+\frac{L^2}{\mu C\left(V_g-V_{th}\right)}\right)^{-1}\,.
\end{equation}
This equation allows for extraction of field effect mobility using a fit to the measured $G$($V_g$). Here, the mobility $\mu$, the interface resistances $R_s$, and the threshold voltage $V_{th}$ are the free fit parameters. We restrict the fitting range to $G^{-1}(V_g) \leq$ 100\,k$\Omega$. We independently calculate the capacitance from a finite element model of the device (see Fig.~\ref{MOB_fig1}c inset), where we take into account that quantum confinement in our nanowires reduces the classical capacitance by $\sim 20 \%$~\cite{Wang2006,Eeltink2013}. Neglecting quantum effects in our capacitance calculation would lower the extracted mobility values by $\sim 20 \%$. Further details on the calculation of the capacitance can be found in Supplementary Text 3. We compared the mobility values extracted by a fit using eq.~\ref{mob_fit_expr} with the mobility values obtained from peak transconductance, a common method to extract nanowire mobility, and found matching results (see Supplementary Text 4). For a representative fabrication run, mean forward mobility of 11 devices is found to be $2.9\mathbin{\times}10^4$\,cm$^2$/Vs using our fit method, whereas peak-transconductance method yields 2.7 (1.9) $\mathbin{\times}10^4$\,cm$^2$/Vs  with (without) taking into account the interface resistances. Our fit method, however, differs from peak transconductance method where the mobility is extracted from the maximum value of the transconductance using a small gate voltage range. Because we consider the transconductance in a wide gate voltage range by fitting a large section of $G$($V_g$), the extracted mobility is insensitive to small conductance fluctuations. This is contrary to the peak transconductance where conductance fluctuations greatly affect the extracted mobility. We show in Supplementary Text 5 that our simple model with gate voltage-independent interface resistances is a valid approximation for our measurements. However, despite our thorough analysis a general drawback of field effect mobility remains: the uncertainty in the calculated capacitance value affects the extracted mobility directly. Nanowires suffer from this drawback as their small dimensions do not allow a straightforward experimental extraction of capacitance.

To determine what limits the mobility in our devices, we systematically studied the effect of various experimental parameters by measuring $\sim$~10 devices simultaneously fabricated on the same substrate. We then change one parameter at a time for each fabrication run to deduce its effect on the field effect mobility.

\section{Results and discussions}

\subsection{Nanowire surface and adsorption}

\begin{figure}[!ht]
	\centering
 	\includegraphics[width=1\textwidth]{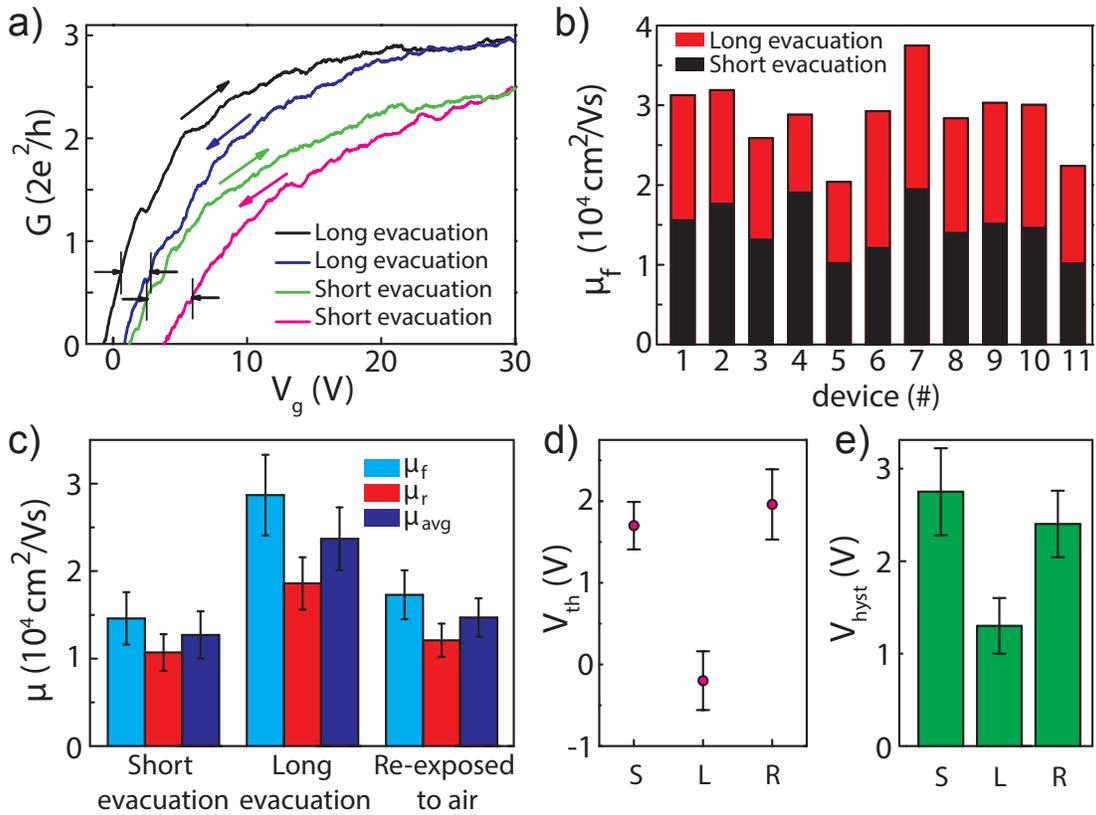}
 	\caption{\textbf{a)} Conductance $G$($V_g$) of samples measured after evacuation of the sample space for a short or long period of time prior to cool down. Samples are evacuated for $\sim$~15 minutes ($\sim$~65 hours), giving the green and pink (black and blue) conductance curves for forward and reverse sweep direction respectively. Arrows indicate sweep direction. The same chip with nanowire devices is first evacuated only shortly (yielding the data denoted with 'short evacuation'), then evacuated for longer-time ('long evacuation' data), reexposed to air for $\sim$~90 hours and evacuated shortly ($\sim$~15 minutes) again ('re-exposure' data), see panel c. The substrate was cleaned prior to nanowire deposition. Hysteresis of both pairs of conductance curves is indicated with arrows and vertical lines. Although the hysteresis is indicated at non-zero $G$, the hysteresis reported in panel c is extracted from the difference in threshold voltage between conductance curves with forward and reverse sweep direction. \textbf{b)} Mobility obtained with forward sweep direction, $\mu_f$, of individual devices after short (black) or long (red) device evacuation time. 
 	\textbf{c)} Mobility after short-time evacuation, long-time evacuation, and reexposure to air. $\mu_{\mathrm{avg}}$ is the average of the mobility obtained with forward sweep direction, $\mu_f$, and with reverse sweep direction, $\mu_r$.
 	\textbf{d)} Threshold voltage extracted from forward sweep direction, $V_{th}$ after short-time evacuation (S), long-time evacuation (L) and reexposure to air (R). 
	\textbf{e)} Hysteresis $V_{\mathrm{hyst}}$, after short-time evacuation (S), long-time evacuation (L) and reexposure to air (R). The hysteresis is given by the difference in threshold voltage between forward and reverse sweep direction. All values in panels c, d and e are an average, obtained from fits to the conductance curve of each device on the measurement chip. Error bars in panels c, d and e indicate the standard deviation.} 
 	\label{MOB_fig2}
\end{figure}

Nanowire conductivity at room temperature is known to increase after evacuation of the sample space following mounting of devices~\cite{Kretinin2010,Nadj-Perge2010a}. We find that evacuation also strongly affects $G$($V_g$) at low temperature (4 K). Comparing the $G$($V_g$) measured for short and long sample space evacuation time prior to cool down, we observe a steeper increase of conductance with gate voltage after long-time evacuation (Fig.~\ref{MOB_fig2}a). Considering a number of devices on the same measurement chip, we find almost a doubling of the mobility values after long-time sample evacuation (Fig.~\ref{MOB_fig2}b). The re-exposure of samples to air after long-time evacuation results in a reduction of mobility (Fig.~\ref{MOB_fig2}c) with values very similar to those obtained from the initial measurements with a short-time sample space evacuation. The transconductance is larger when the gate is swept from low towards high voltages (forward sweep direction) leading to higher mobility compared to the case of sweeping from high gate voltages to low (reverse sweep direction) (Fig.~\ref{MOB_fig2}c). Moreover, after long-time evacuation a shift of the threshold voltage towards more negative values is observed (Fig.~\ref{MOB_fig2}d) together with a reduced hysteresis (Fig.~\ref{MOB_fig2}e). Both the threshold voltage and the hysteresis regain their initial values obtained from short-time evacuation once the sample is re-exposed to air, similar to the extracted mobility: exposing the devices to air has a reversible effect on the field effect transport parameters we extract from the fits. All extracted fit parameters can be found in Supplementary Table 1. 

A hysteresis in transconductance dependent on ambient conditions has been studied before by Kim \textit{et al}~\cite{Kim2003} and Wang \textit{et al}~\cite{Wang2004}, and was attributed to the adsorption of water onto the nanostructure and onto the SiO$_2$ substrate. Evacuation of the sample environment leads to desorption of water, thereby reducing the hysteresis. However sample evacuation alone is insufficient to fully remove the adsorbed water. The similarities between our observations and those reported by Wang \textit{et al} and Kim \textit{et al}, considering both the influence of gate voltage sweep direction on the shift of the threshold voltage, as well as the reduction of hysteresis with evacuation time and the reversibility of the effect when reexposing samples to air, strongly suggest that the field effect transport is affected by molecules adsorbed to the nanowire and/or the SiO$_2$ substrate. Water is highly likely to be the main adsorbate because reexposing the device to ambient atmosphere following long evacuation time of sample space yields values of mobility, threshold voltage and hysteresis similar to those obtained from the measurements with short evacuation time. InSb nanowires have however also shown decreased conductance in response to isopropanol and acetone~\cite{Penchev2012}. 

It is an open question how adsorbates affect device conductance at low temperature. The alignment of polar molecules by gate electric field may result in an additional gating~\cite{Wang2004}. However, the mechanism through which such alignment causes hysteresis is not clear. Another scenario is charge trapping by adsorbed molecules~\cite{Kim2003}. Such trapping could possibly lead to an asymmetry between forward and reverse sweep direction, yielding the observed hysteresis and sweep direction dependent mobility. The observed trapping mechanism is likely to have a long response time, as our measurements are taken at relatively low gate voltage sweep rates (120\,mV/s). Unlike refs.~\cite{Dayeh2007,Kim2003,Wang2004}, we find no dependence on sweep rate for rates between 3 -- 600\,mV/s. Nonetheless, repeated measurements yield the same $G(V_g)$, implying that between scans the traps are emptied.

\subsection{Substrate cleaning}

\begin{figure}[!ht]
	\centering
 	\includegraphics[width=1\textwidth]{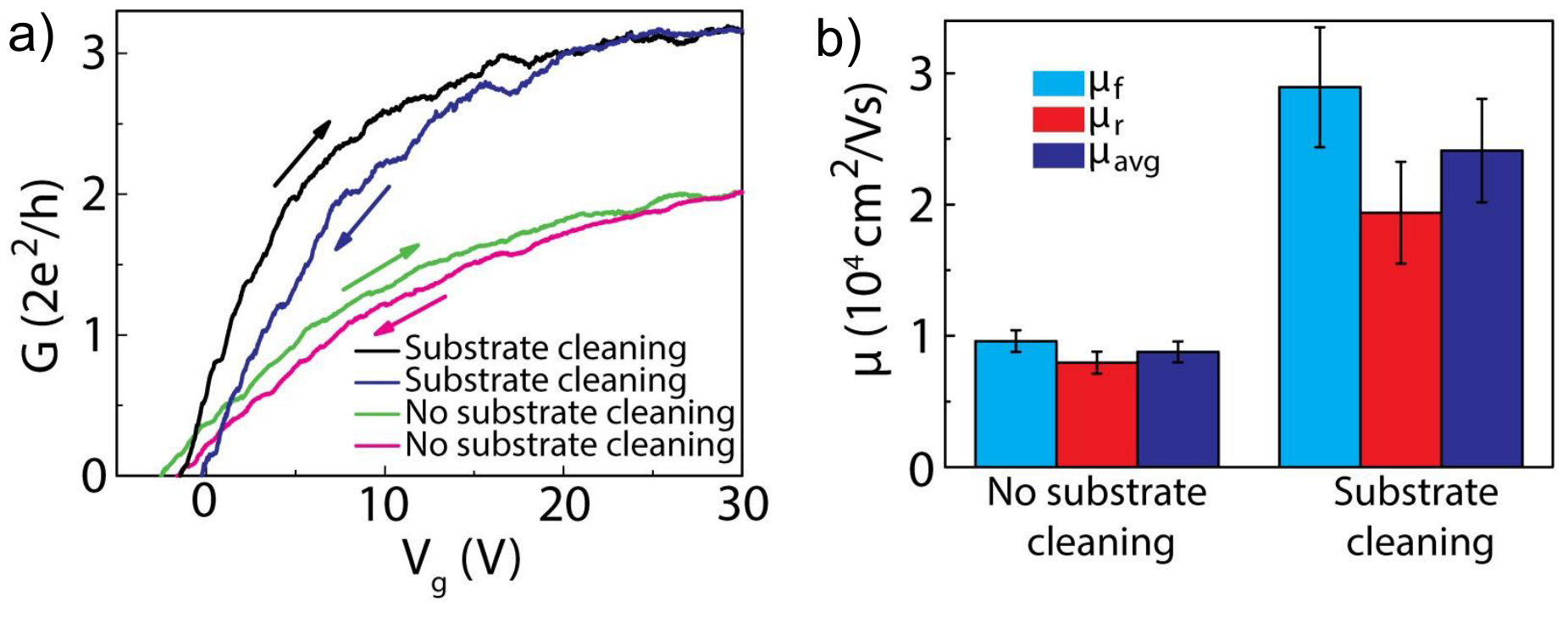}
 	\caption{\textbf{a)} Conductance curves $G(V_g)$ obtained from samples without and with substrate cleaning. Forward and reverse sweep direction are indicated with arrows. Samples have been evacuated for $\sim$~60 hours before cool down. \textbf{b)} Forward, reverse and average mobility with and without substrate cleaning. Values are averages obtained from fits to conductance curves of individual devices. Error bars indicate standard deviation.}
 	\label{MOB_fig3}
\end{figure}

We further find that cleaning of Si/SiO$_2$ substrates by remote oxygen plasma prior to nanowire deposition results in an enhanced gate dependence of low-temperature conductivity. Fig.~\ref{MOB_fig3}a shows $G$($V_g$) curves of individual devices, while Fig.~\ref{MOB_fig3}b shows an average over extracted mobilities obtained from measurements of $\sim$~10 FETs with and without substrate cleaning. All other fabrication and measurement steps are the same for both sets of devices. The remote oxygen plasma most probably removes hydrocarbons that remain on the substrates after fabrication of alignment markers or during storage of samples in a polymer-containing environment. We verified that the oxygen plasma cleaning does not decrease the thickness of the SiO$_2$ gate dielectric within the measurable range $<1$\,nm.

\subsection{Contact spacing}

\begin{figure}[!ht]
	\centering
 	\includegraphics[width=0.6\textwidth]{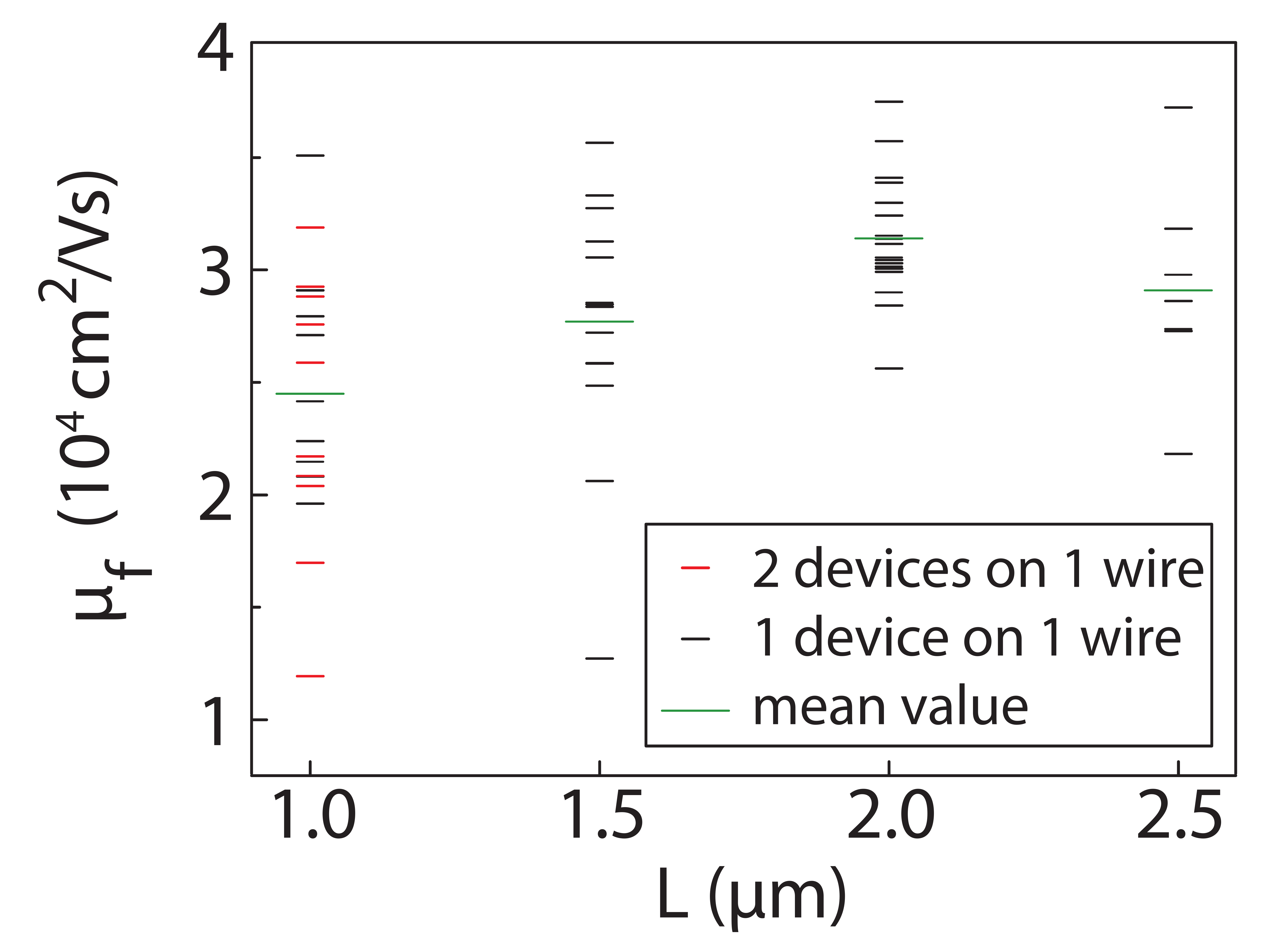}
 	\caption{Mobility obtained by sweeping the gate voltage in forward direction, $\mu_f$, as a function of source-drain contact spacing $L$. Data from 5 different measurement chips (see Supplementary Text 6). Red lines indicate mobility values obtained from long nanowires on which three contact electrodes were placed, resulting in two FETs in series, while black lines correspond to the mobility of single FET devices. Mean forward mobility for each contact spacing is $\mu_{f,m}(L = 1$\,$\mu$m) = 2.4$\times$10$^4$\,cm$^2$/Vs, $\mu_{f,m}(L = 1.5$\,$\mu$m) = 2.8$\times$10$^4$\,cm$^2$/Vs, $\mu_{f,m}(L = 2$\,$\mu$m) = 3.1$\times$10$^4$\,cm$^2$/Vs and $\mu_{f,m}(L = 2.5$\,$\mu$m) = 2.9$\times$10$^4$\,cm$^2$/Vs.}
 	\label{MOB_fig4}
\end{figure}

A correlation between FET source-drain contact spacing and extracted field effect mobility is found (Fig.~\ref{MOB_fig4}). Although the spread in mobility at a given contact spacing is substantial, an overall increase of extracted mobility is observed with increasing contact spacing. To determine whether the dependence of the field effect mobility on contact spacing originates from the length of the used nanowire, FETs with short (1\,$\mu$m) contact spacing were realized both on short wires, and on long wires using three contact electrodes resulting in two FETs in series. Devices made from both long and short wires with 1\,$\mu$m contact spacing give similar mobility (see Fig.~\ref{MOB_fig4}). The contact spacing dependence is thus a device property rather than a nanowire property.

A reduced mobility for short contact spacing is expected when transport is \mbox{(quasi-)}\mbox{ballistic} rather than diffusive~\cite{Shur2002,Chen2008}. We have observed ballistic transport in our wires~\cite{vanWeperen2013} with a device geometry and measurement conditions different from those here. Here we expect quasi-ballistic transport in our devices with a mean free path comparable to nanowire diameter $l_e\sim$~0.1\,$\mu$m. While devices with $L/l_e \gg 1$ are preferable, our InSb nanowires can currently not be grown longer than $\sim$~3.5\,$\mu$m. However, while for channel length of 1\,$\mu$m \mbox{(quasi-)}\mbox{ballistic} effects may play a role, mobility values obtained from our devices with longer contact spacing yield a better estimate of field effect mobility. Moreover, effects related to the metal contacts are expected to play a larger role in devices with short contact spacing and can possibly contribute to the observed decrease of $\mu$($L$) in short channel devices. Possible explanations are that (1) the contacts reduce the capacitance of short devices more than expected from the Laplace simulations (in which the nanowire is assumed to be metallic) or (2) electrons are injected from and absorbed over a finite length underneath the contacts, leading to an effective $L$ larger than the contact spacing.

\subsection{Reproducibility}

\begin{figure}[!ht]
	\centering
 	\includegraphics[width=0.7\textwidth]{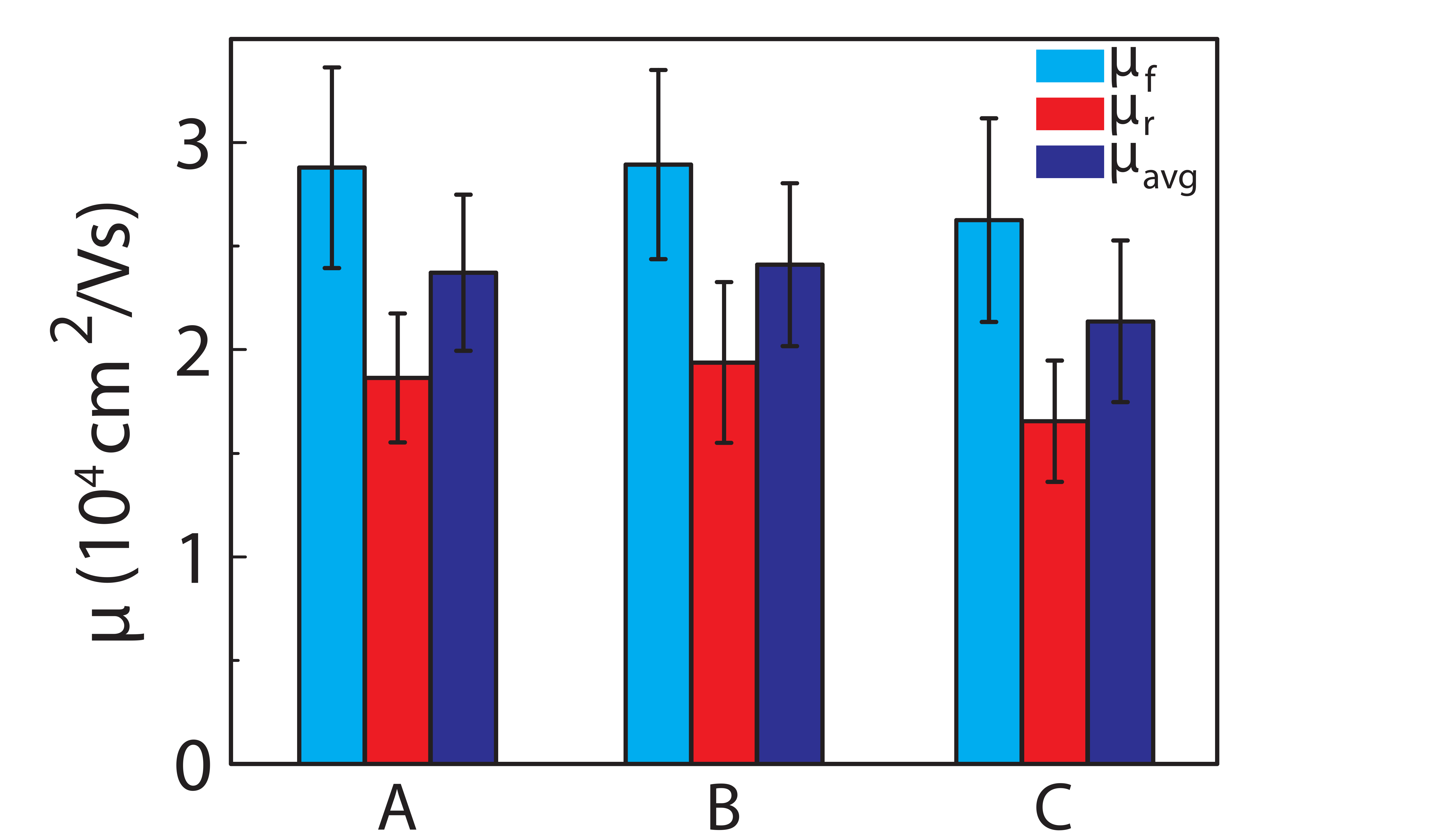}
 	\caption{Average mobilities obtained with forward ($\mu_f$) and reverse ($\mu_r$) sweep direction. First group of data (batch A) corresponds to the fabrication run presented in Fig.~\ref{MOB_fig2} (long-time evacuation), batch B is presented in Fig.~\ref{MOB_fig3}b (with substrate cleaning), whereas batch C is a separate batch to demonstrate the reproducibility of our results. Average mobility $\mu_{\mathrm{avg}}$ is the average of forward and reverse mobility. All results are obtained by improved cleaning of the substrate and long evacuation time of the sample space. Error bars indicate standard deviation.}
 	\label{MOB_fig5}
\end{figure}

Altogether, cleaning the SiO$_2$ substrate before wire deposition and applying a long sample evacuation time yields $\mu_{avg} \approx 2.5\mathbin{\times}10^4$\,cm$^2$/Vs for devices with a contact spacing $L = 2$\,$\mu$m. This mobility is the average value of $\mu_f = 3.1\mathbin{\times}10^4$\,cm$^2$/Vs (see Fig.~\ref{MOB_fig4}) and $\mu_r = 1.9\mathbin{\times}10^4$\,cm$^2$/Vs. These high mobilities result from measurements of $\sim$~15 devices fabricated in different fabrication runs (see Supplementary Text 6 for details) using the same fabrication recipe. Fig.~\ref{MOB_fig5} demonstrates the reproducibility of our results: mobility obtained from three different fabrication runs are very similar. The optimized nanofabrication recipe as well as an overview of all the parameters extracted from the fits to the conductance vs. gate voltage curves that yield Fig.~\ref{MOB_fig5} are given in Supplementary Text 1 and Supplementary Table 2, respectively. 

\section{Conclusions and outlook}

Low-temperature field effect mobility of InSb nanowires is extracted by measuring the conductance as a function of gate voltage. Taking surface adsorption and substrate cleaning into consideration, an optimized nanofabrication recipe has been obtained yielding average field effect mobilities of $\sim 2.5\mathbin{\times}10^4$\,cm$^2$/Vs. It is demonstrated that the obtained mobility values are highly reproducible. 

As we show that surface adsorption has a large impact on field effect mobility, further studies should be directed towards minimizing the adsorbates and analysis of surface properties. An improved design of the measurement setup allowing for heating and better evacuation of the sample space is likely to facilitate a further desorption of adsorbates. Exposing the devices to UV-light during evacuation, which may assist desorption, can also be investigated~\cite{Penchev2012}. Further, passivating the nanowire surface by removing the native oxide followed by application of a high quality dielectric likely reduces surface adsorption. Possible methods are atomic hydrogen cleaning~\cite{Hjort2014} or chemical etching followed by dielectric deposition~\cite{Hou2008}. Alternatively, by suspending the nanowires above a metallic gate using vacuum as a dielectric, one can minimize the effects of the substrate adsorption, leaving the wire adsorption as the predominant constituent affecting the field effect mobility. In the case of adsorbates creating a fluctuating potential profile along the wire resulting in charge scattering, a core-shell structure is expected to yield a higher field effect mobility because the potential fluctuations due to adsorbates are spatially separated from the channel owing to the shell. Finally, to study the surface composition of the nanowire and the substrate, x-ray photoelectron spectroscopy or Auger electron spectroscopy could be used~\cite{Chu2011}.

\section*{Acknowledgement}
We thank K.~Zuo, D.~Szombati, V.~Mourik,  A.~Geresdi and J.~W.~G. van den Berg for preliminary studies and fruitful discussions. This work has been supported by Dutch Organisation for Scientific Research (NWO), Foundation for Fundamental Research on Matter (FOM), European Union Seventh Framework Programme under grant agreement no. 265073 (NANOWIRING), and Microsoft Corporation Station Q.

\end{document}


\title{Supplementary Part: Towards high mobility InSb nanowire devices}

\author{\"Onder G\"ul$^{1*}$, David J.~van Woerkom$^{1*}$, Ilse van Weperen$^{1*}$, Diana Car$^2$, S\'ebastien R.~Plissard$^{1,2,3}$, Erik P.~A.~M.~Bakkers$^{1,2}$, Leo P.~Kouwenhoven$^1$}
\address{$^1$QuTech and Kavli Institute of Nanoscience, Delft University of Technology, 2600 GA Delft, Netherlands}
\address{$^2$Department of Applied Physics, Eindhoven University of Technology, 5600 MB Eindhoven, Netherlands}
\address{$^3$Present address: Laboratoire d'Analyse et d'Architecture des Syst\`emes, 7, avenue du Colonel Roche BP 54200 31031 Toulouse cedex 4, France}
\address{*Equal contribution}
\ead{\mailto{o.gul@tudelft.nl}}

\section{Optimized fabrication recipe}\label{mob_recipe}

\begin{itemize}

\item Substrate cleaning: 10 minutes remote oxygen plasma cleaning (Tepla 300 Plasma Asher) of the p$^{++}$-Si substrate covered with 285\,nm dry thermal SiO$_{2}$ with pre-defined Au alignment markers (oxygen pressure 1\,mbar, plasma power 600\,W). All substrates were from the same wafer.

\item Wire deposition: deterministic positioning of wires using a setup similar to that described in ref.~[35]. Wires were always taken from the same section on the same growth chip.

\item SEM imaging of the nanowires with surrounding alignment markers. Images are used for the subsequent design of the contacts.

\item Spin resist: PMMA 950A4 at 4\,krpm, baking $>$ 15 minutes at a temperature of 175\,$^\circ$C.

\item Electron beam writing of the contact design.

\item Development: MIBK:IPA 1:3 60\,s, IPA 60\,s.

\item Ar etching (AJA International sputtering system) using rf-plasma: pressure 3\,mTorr, Ar flow 50\,sccm, power 100\,W, duration 300\,s, no rotation of the sample holder. A voltage of 300\,V is applied to the sample holder.

\item Contact deposition: e-beam evaporation of Ti/Au 5/145\,nm with deposition rate 0.5\,\r{A}/s and 2\,\r{A}/s respectively.

\item Lift-off in acetone: the sample with acetone is heated for several hours and left in acetone for $\geq$ 12\,h.

\item Samples were stored in an Ar glove box between fabrication and mounting. 

\end{itemize}

\section{Measurements}\label{measurement}

\begin{itemize}

\item Sample space (IVC) evacuated for $\sim$ 60 hours after mounting (insert type: Desert Cryogenics). 

\item For thermalisation, He of $\sim$10\,mbar is added to sample space at room temperature before cooling down the devices. During low-temperature measurements samples are kept in a vacuum environment. 

\item $G$($V_g$) measured using 10\,mV bias, gate voltage range from $-$6\,V to +30\,V with sweep rate 6\,mV/50\,ms. Measured in forward and reverse sweep direction.

\item To check a possible sweep rate dependence of $G$($V_g$), gate voltage steps of 0.15, 0.3, 0.6, 1.5, 3, 6, 15, 30 [mV/(50\,ms)] is used both in forward and reverse sweep direction. No dependence on sweep rate was found.

\end{itemize}

\section{Device capacitance}\label{capacitance}

The capacitance for different channel lengths is calculated with a 3D Laplace solver for a realistic device geometry including the metallic leads. Here the wire is assumed to be metallic. Then, for a more accurate representation of the device capacitance and to account for quantum confinement effects, 2D Schr\"odinger-Poisson solver was used and its result is compared with the capacitance calculated with 2D Laplace solver. A reduction of capacitance by 20\% is found for the case of quantum mechanical treatment of the wire. In Fig. 1c (inset) in the main text, the capacitance calculated by 3D Laplace solver with a 20\% reduction is plotted. These plotted capacitance values are used for mobility extraction and expected to represent channel capacitance realistically.

\clearpage \newpage

\section{Comparison of field effect mobility extraction methods}\label{comparison of methods}

\begin{figure}[p!]
	\centering
 	\includegraphics[width=0.75\textwidth]{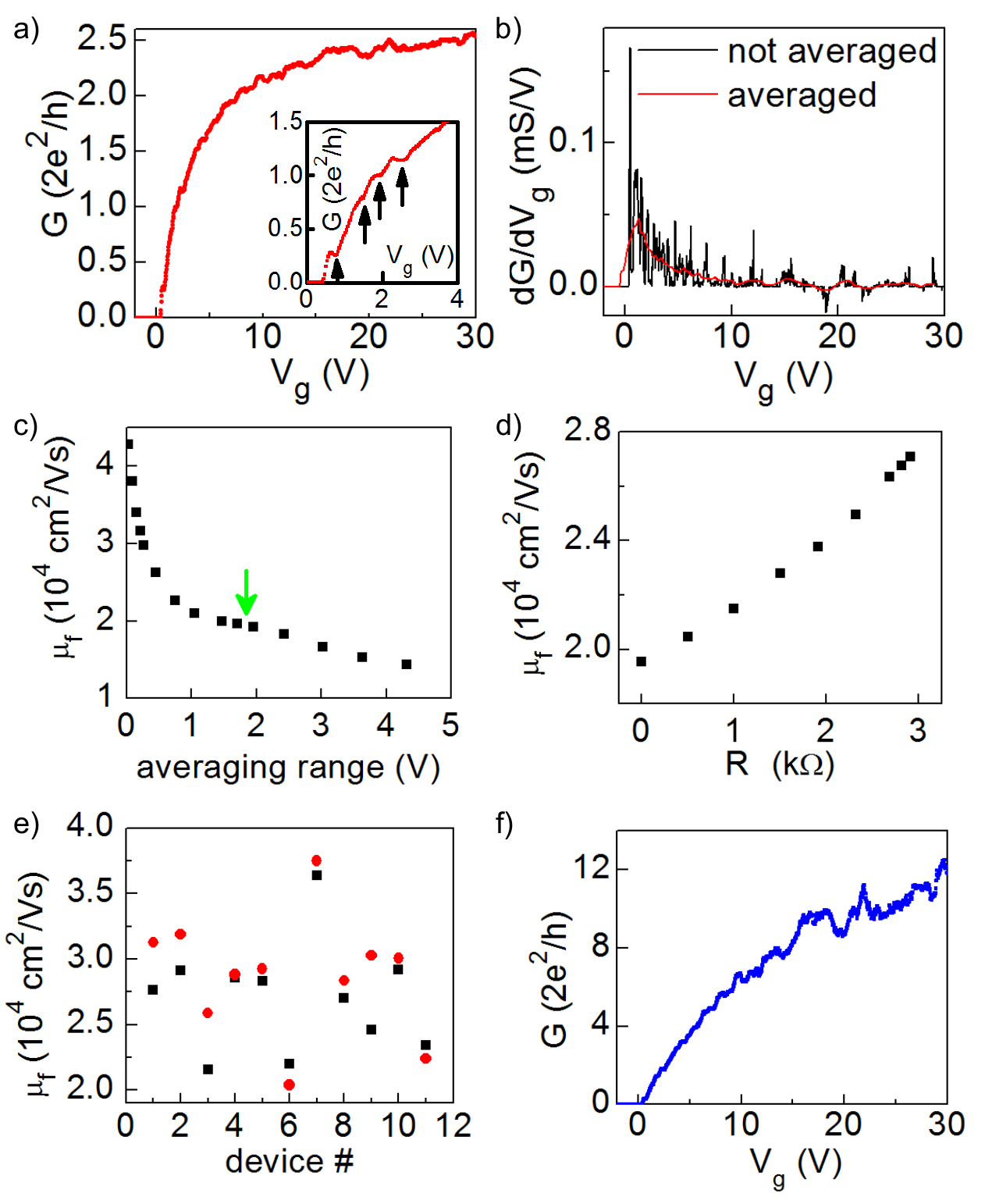}
 	\caption{\textbf{a)} Conductance $G$, as a function of gate voltage $V_g$. Inset: Zoom-in of the conductance near pinch-off. The arrows point at universal conductance fluctuations resulting in fluctuations in transconductance. \textbf{b)} Transconductance $dG/dV_g$ without (black) and with (red) averaging over 1.8\,V gate voltage range. Averaging is applied to remove the fluctuations that lead to peaks and dips in the transconductance. \textbf{c)} Field effect mobility $\mu_{f}$ obtained from peak transconductance as a function of gate voltage averaging range. Plotted values of peak-mobility is the average of 11 devices on the same chip (long evacuation time experiment, Fig.~2). The green arrow denotes the averaging range of 1.8\,V used for the averaged curve in panel b. This averaging window is used in further analysis to obtain peak-mobility. \textbf{d)} Peak-mobility as a function of series resistance subtracted from $G(V_g)$. Peak-mobility is the average of 11 devices on the same chip. \textbf{e)} Comparison between field effect mobility $\mu_{f}$ obtained for individual devices using the fit according to eq.~3 (red points) and the mobility obtained from peak transconductance (black points). \textbf{f)} Conductance as a function of gate voltage after the correction for interface resistances. For this device an interface resistance $R_s$ = 4 k$\Omega$ is assumed. Conductance curve without the correction for $R_s$ is shown in panel a.}
 	\label{MOB_figS1}
\end{figure}

We extract mobility values by fitting the conductance curves $G$($V_g$) in a large gate voltage range. However, in the literature mobility is commonly extracted from a small gate voltage range where the transconductance has its maximum value (peak transconductance). This gate voltage range is typically close to the threshold voltage where the mobility is expected to be the highest. Here, we compare the field effect mobility obtained using our method -- fitting the conductance curves $G$($V_g$) -- to the field effect mobility obtained from peak transconductance, the standard method for extracting mobility in nanowires. We denote the mobility obtained using the latter as peak-mobility. We describe the extraction of peak-mobility in the following: By numerically differentiating the measured $G$($V_g$) shown in Fig.~\ref{MOB_figS1}a, one obtains the transconductance $g_m=dG/dV_g$. This transconductance is shown in Fig.~\ref{MOB_figS1}b (black curve). After taking the numerical derivative, an averaging is performed to remove the fluctuations in transconductance (red curve in Fig.~\ref{MOB_figS1}b). The peak-mobility is then obtained from the maximum value of transconductance using $\mu = g_m L^2/C$ (see eq.~2). Peak-mobility depends strongly on the chosen averaging range. This dependence is shown in Fig.~\ref{MOB_figS1}c. Here, mean forward peak-mobility of 11 devices from a single fabrication run is plotted against the averaging range. We choose the averaging range to be 1.8\,V, the value at which the rapid decrease of peak-mobility with respect to averaging range diminishes.

Next interface resistances are taken into account since they affect the extracted peak-mobility. This is done by subtracting the contribution of a gate-independent series resistance $R$ from the measured conductance curve $G$($V_g$). Fig.~\ref{MOB_figS1}f shows an example of such a conductance curve corrected for interface resistances. From such a curve we determine the transconductance, and from the maximum value of transconductance peak-mobility is extracted. The peak-mobility depends on the subtracted $R$, shown in Fig.~\ref{MOB_figS1}d. Here, as mentioned above, mean forward peak-mobility of 11 devices from a single fabrication run is plotted. (The peak-mobility for $R = 0$ is the one indicated with a green arrow in Fig.~\ref{MOB_figS1}c.) For zero subtracted resistance ($R = 0$), the transconductance has a global maximum near pinch-off (Fig.~\ref{MOB_figS1}b, red curve). Upon increasing the value of $R$ subtracted from $G$($V_g$), the transconductance values increase for all gate voltages, with the amount of increase being larger for higher gate voltages. When $R$ exceeds the value of interface resistances $R_s$, the transconductance no longer has a global maximum near pinch-off. When $R$ is increased even further, transconductance starts to increase with gate voltage, a case we regard to be unrealistic. $R_s$ for individual devices varies between 1.5\,k$\Omega$ and 4\,k$\Omega$, with an average $R_s$ of $\sim$~3\,k$\Omega$. After the subtraction of $R_s$, the mean peak-mobility of 11 devices obtained using forward sweep direction is (27.1 $\pm$ 4.2) $\times$ 10$^3$ cm$^2$/Vs (see Fig.~\ref{MOB_figS1}d) compared to (28.7 $\pm$ 4.8)$\times$ 10$^3$ cm$^2$/Vs obtained from fits to the conductance curves. Both values are within error margin the same. Comparing mobilities of individual devices obtained using both methods (Fig.~\ref{MOB_figS1}e), we conclude that both methods give similar values. The small difference is due to slightly larger interface resistances obtained from the fitting method, giving an average $R_s$ of 3.7\,k$\Omega$.

\section{Simplification of gate voltage-independent interface resistances}\label{interface resistance}

\begin{figure}[ht]
	\centering
 	\includegraphics[width=1\textwidth]{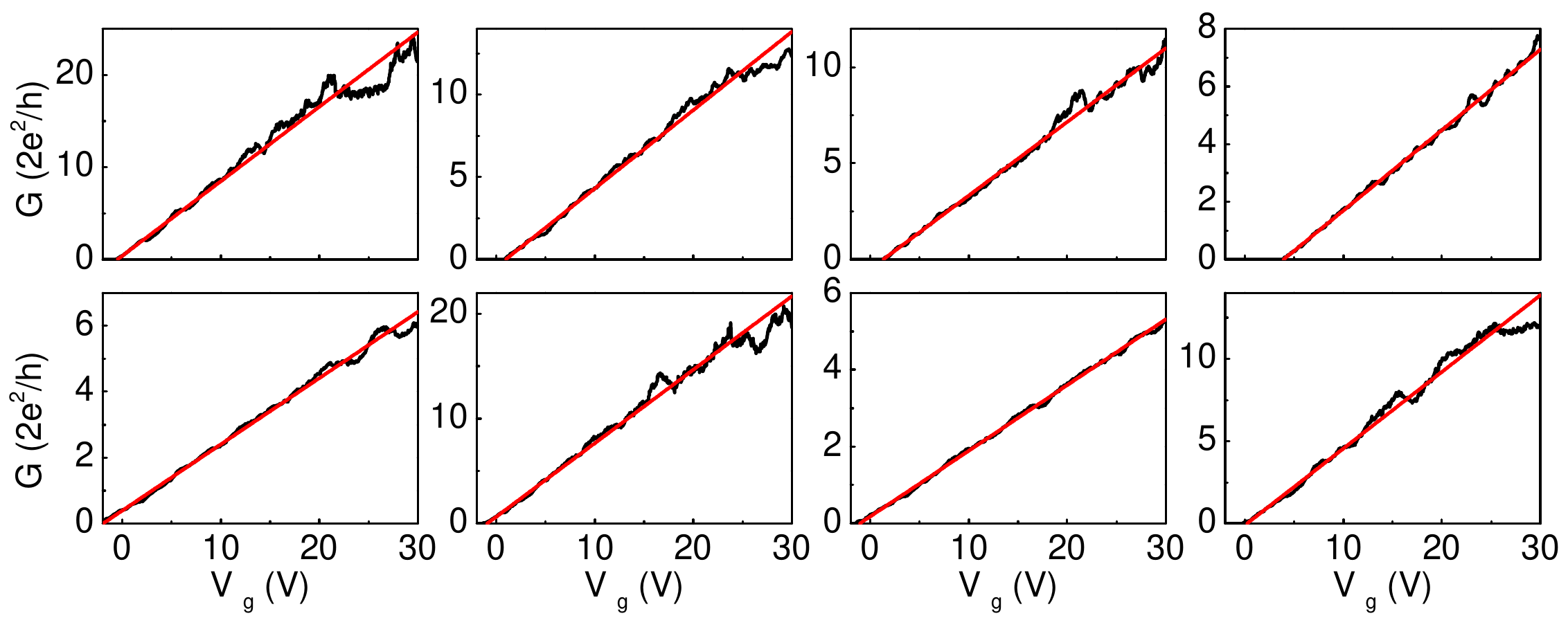}
 	\caption{Panels show the measured device conductance after subtracting the interface resistance, $G_{\mathrm{ch}}(V_g)$ (black), together with $G_{\mathrm{L}}(V_g)$ (red), which is the conductance linear in gate voltage with the transport properties extracted from the fit. $G_{\mathrm{ch}}(V_g)$ shown in upper row (lower row) are from the data set presented in Fig.~2a (Fig.~3a) in the main text.}
 	\label{MOB_figS2}
\end{figure}

Here we check our simplification of modelling the interface resistances $R_s$ to be gate voltage independent. We fit the measured device conductance $G(V_g)$ using eq.~3 to determine $R_s$, the mobility $\mu$, and the threshold voltage $V_{th}$. The measured device conductance after the subtraction of $R_s$ is denoted by $G_{\mathrm{ch}}(V_g)$. In our model $G_{\mathrm{ch}}(V_g)$ has the form $G_{\mathrm{L}}(V_g)=(V_g-V_{th})\mu C/L^2$, which corresponds to a conductance linear in gate voltage with the transport properties extracted from the fit mentioned above. In Fig.~\ref{MOB_figS2} we plot representative curves of $G_{\mathrm{ch}}(V_g)$ (black) and compare them with $G_{\mathrm{L}}(V_g)$ (red). We find that $G_{\mathrm{L}}(V_g)$ matches well with $G_{\mathrm{ch}}(V_g)$, demonstrating that our simple model with gate voltage-independent interface resistances is a valid approximation for our measurements.

\clearpage \newpage

\section{Overview of measured devices}\label{overview text}

\begin{itemize}

\item Evacuation time experiment (fabricated according to our optimized recipe described above). Mobility was extracted from measurements of 11 devices with contact spacing between 1 and 2\,$\mu$m. Average contact spacing 1.41\,$\mu$m. Data reported in Fig.~2. Long evacuation time data is also included in Fig.~4 and Fig.~5 (Batch A).

\item FETs without substrate cleaning and with long-time evacuation.  Fabricated according to our optimized recipe, with the exception that we used different settings for Ar etching. Here we used 400\,V on the sample holder and etched for 150\,s while keeping all the other settings the same. This yields the same amount of etching of InSb nanowire ($\sim$~70\,nm) as etching at 300\,V for 300\,s. 11 devices, contact spacing of all devices is 2\,$\mu$m. Data reported in Fig.~3. 

\item FETs with substrate cleaning and with long-time evacuation. Fabricated according to our optimized recipe, with the exception that we used different settings for Ar etching. Here we used 400\,V on the sample holder and etched for 150\,s while keeping all the other settings the same. This yields the same amount of etching of InSb nanowire ($\sim$~70\,nm) as etching at 300\,V for 300\,s. 11 devices, contact spacing between 1 and 2\,$\mu$m. Average length 1.42\,$\mu$m. Data reported in Fig.~3, Fig.~4, and Fig.~5 (Batch B).

\item FETs fabricated according to our optimized recipe. 13 devices, contact spacing between 1 and 2.5\,$\mu$m. Average contact spacing 1.73\,$\mu$m. Data reported in Fig.~4 and Fig.~5 (Batch C).

\item FETs fabricated according to the recipe, with the addition of a thin layer of perfluorodecyltrichlorosilane (FDTS) deposited onto the devices after fabrication. No improvement of mobility was observed with respect to devices without FDTS. 11 devices, contact spacing between 1 and 2.5\,$\mu$m. Average contact spacing 1.64\,$\mu$m. Data reported in Fig.~4.

\item FETs fabricated according to our optimized recipe, but with substrate oxygen plasma cleaning of 60\,s instead of 10 minutes. After oxygen plasma cleaning a thin layer of FDTS was deposited onto the substrate, after which fabrication proceeded according to the recipe. No improvement of mobility was observed with respect to devices without FDTS and with the usual 10 minutes cleaning. 10 devices, contact spacing between 1 and 2.5\,$\mu$m. Average contact spacing 1.85\,$\mu$m. Data reported in Fig.~4.

\end{itemize}

\clearpage \newpage

\section{Average device characteristics obtained from several measurement and fabrication runs}\label{overview table}

\begin{table}[h!]  
	\centering
       \begin{tabular}{| l || c | c | c | }
        \hline
                                                       & Short-time evacuation         &          Long-time evacuation     &   Reexposed to air
        \\
    \hline \hline
    $\mu_f$ (10$^3$ cm$^2$/Vs)         &  14.6 $\pm$ 3.0    &       28.7 $\pm$ 4.6   &   17.3 $\pm$ 2.8          \\ \hline
    $\mu_r$ (10$^3$ cm$^2$/Vs)         &  10.7 $\pm$ 2.1     &       18.6 $\pm$ 3.0   &   12.1 $\pm$ 1.9          \\ \hline
    $\mu_{\mathrm{avg}}$ (10$^3$ cm$^2$/Vs)         &  12.7 $\pm$ 2.7    &       23.7 $\pm$ 3.6   &   14.7 $\pm$ 2.2          \\ \hline
    $V_{th}$     (V)                           &   1.70 $\pm$ 0.29    &       -0.20 $\pm$ 0.36   &   1.96 $\pm$ 0.43        \\ \hline
    $V_{\mathrm{hyst}}$  (V)                              &   2.75 $\pm$ 0.47    &       1.31 $\pm$ 0.30    &   2.40 $\pm$ 0.36          \\ \hline
    $R_{s}$ ($k\Omega$)                       &   3.7 $\pm$ 0.7     &       3.7 $\pm$ 1.0     &   4.1 $\pm$ 1.2         \\ \hline
    \end{tabular}
    \caption{Mobility, threshold voltages $V_{th}$, hysteresis, $V_{\mathrm{hyst}}$ and series resistances, $R_{s}$, extracted from fits to conductance curves $G$($V_g$) of the evacuation time experiment. Mobility is obtained with forward sweep direction, $\mu_f$ and reverse sweep direction, $\mu_r$. The average mobility of these two sweep directions, $\mu_{\mathrm{avg}}$, is also reported. $V_{th}$ is the threshold voltage obtained from fits to $G$($V_g$) taken with forward sweep direction. Mobility, threshold voltage and hysteresis are also shown in Fig.~2c, d and e, respectively.}
    \label{table_evacuation}
\end{table}

\begin{table}[h!] 
	\centering
    \begin{tabular}{| l || c | c | c |}
        \hline
                                      Batch       & A          				    &          B       					&   C                      \\
    \hline \hline

    $\mu_f$   (10$^3$ cm$^2$/Vs)									&  28.7 $\pm$ 4.6       &          28.9  $\pm$ 4.4   &   26.0 $\pm$ 4.7            \\ \hline
    $\mu_r$   (10$^3$ cm$^2$/Vs) 									&  18.6 $\pm$ 3.0       &          19.4  $\pm$ 3.9   &   16.4 $\pm$ 3.0            \\ \hline
    $\mu_{\mathrm{avg}}$      (10$^3$ cm$^2$/Vs) 	&  23.7 $\pm$ 3.6       &          24.2  $\pm$ 3.9   &   21.2 $\pm$ 3.8            \\ \hline
    $V_{th}$     (V)                    					& -0.20 $\pm$ 0.36      &         -0.51 $\pm$ 0.45   &   -0.37 $\pm$ 0.39           \\ \hline
    $V_{\mathrm{hyst}}$  (V)                			&  1.31 $\pm$ 0.30      &          1.14  $\pm$ 0.22  &   1.41 $\pm$ 0.28           \\ \hline
    $R_{s}$       ($k\Omega$)                			&   3.7 $\pm$ 1.0       &           3.0   $\pm$ 0.8  &   4.8 $\pm$ 1.8            \\ \hline
    \end{tabular}
    \caption{Mobility, threshold voltage, $V_{th}$, hysteresis, $V_{hyst}$, and series resistance, $R_s$, obtained from fits to the conductance curves $G$($V_g$) of three batches of high-mobility devices. Mobility is obtained with forward sweep direction, $\mu_f$ and reverse sweep direction, $\mu_r$. The average mobility of these two sweep directions, $\mu_{avg}$, is also reported. $V_{th}$ is the threshold voltage obtained from fits to $G$($V_g$) taken using forward sweep direction. Mobilities and series resistances are also shown in Fig.~5.}
    \label{table_reproducibility}
\end{table}